\documentclass[prl,nofootinbib,preprint,superscriptaddress]{revtex4}

\usepackage{graphicx}
\usepackage{amsmath}
\usepackage{amsfonts}
\usepackage{amssymb}
\usepackage{dsfont}
\usepackage{dcolumn}
\usepackage{bm}
\usepackage{pstricks}
\usepackage{color}

\newcommand{\beqn}{\begin{eqnarray}}
\newcommand{\eeqn}{\end{eqnarray}}
\newcommand{\be}{\begin{equation}}
\newcommand{\ee}{\end{equation}}

\newcommand{\mathsym}[1]{{}}

\begin{document}

\title{Developments in Supergravity Unified Models\footnote{To appear in ``Perspectives on
Supersymmetry II", Edited by Gordon Kane, World Scientific, Singapore.}}
\author{Richard Arnowitt }
\affiliation{George P. and Cynthia W. Mitchell Institute for Fundamental Physics and Astronomy, Department of Physics, Texas A\&M University
 College Station, TX  77843-4242}
 \author{Pran Nath}
\affiliation{Department of Physics, Northeastern University,
 Boston, MA 02115, USA}




\begin{abstract}
A review is given of  developments in supergravity unified models proposed in 1982  and their implications 
 for current and future experiment are discussed. \end{abstract}

\maketitle

\section{I. Introduction}\label{sec1}
Supersymmetry (SUSY) was initially introduced as a global symmetry \cite{pramond,golf} on
 purely theoretical grounds that nature should be symmetric between bosons and
 fermions.  It was soon discovered, however, that models of this type had a
 number of remarkable properties \cite{bzum}.  Thus the bose-fermi symmetry led to the
 cancellation of a number of the infinities of conventional field theories, in
 particular the quadratic divergences in the scalar Higgs sector of the Standard
 Model (SM).  Thus SUSY could resolve the gauge hierarchy problem that plagued
 the SM.  Further, the hierarchy problem associated with grand unified models \cite{hgeo}
  (GUT), where without SUSY, loop corrections gave all particles GUT size
 masses \cite{egil,sdim} was also resolved.  
  In addition, SUSY GUT models with minimal
 particle spectrum raised the value for the scale of grand unification,
 $M_G$, to
 $M_G\cong 2\times 10^{16}$ GeV, so that the predicted proton decay rate 
 \cite{sdim,srab}
 was consistent with existing experimental bounds.  Thus in spite of the lack of
 any direct experimental evidence for the existence of SUSY particles,
 supersymmetry became a highly active field among particle theorists.

However, by about 1980, it became apparent that global supersymmetry was
 unsatisfactory in that a phenomenologically acceptable picture of spontaneous
 breaking of supersymmetry did not exist.  Thus the success of the SUSY grand
 unification program discussed above was in a sense spurious in that the needed
 SUSY threshold $M_S$ (below which the SM held) could not be theoretically
 constructed.  In order to get a phenomenologically viable model, one needed
  ``soft breaking'' masses \cite{lgir} (i.e. supersymmetry breaking terms of dimenison
 $\leq$ 3 which maintain the gauge hierarchy)  and these had to be
 introduced in an ad hoc fashion by hand.  In the minimal SUSY model, the
 MSSM \cite{fora},
 where the particle spectrum is just that of the supersymmeterized SM, one could
 introduce as many as 105 additional parameters (62 new masses and mixing angles
 and 43 new phases) leaving one with a theory with little predictive power.
 
 A resolution of the problem of how to break  supersymmetry spontaneously was
 achieved by promoting supersymmetry to a local symmetry\cite{Nath:1975nj},
 and specifically supergravity \cite{dzfr}.
 Here gravity is included into the dynamics.  One can then construct
 supergravity [SUGRA] grand unified models \cite{chams,applied} where the spontaneous
 breaking of supersymmetry occurs in a ``hidden'' sector via supergravity
 interactions in a fashion that maintains the gauge hierarchy.  In such theories
 there remains, however, the question of at what scale does supersymmetry break,
 and what is the ``messenger'' that communicates this breaking from the hidden
 to the physical sector.  In this chapter we consider models where supersymmetry
 breaks at a scale $Q > M_G$ with gravity being the 
 messenger\cite{chams,Barbieri:1982eh,Hall:1983iz,Nath:1983aw,alte}. 
  Such models
 are economical in that both the messenger field and the agency of supersymmetry
 breaking are supersymmetrized versions of  fields and interactions that already
 exist in nature (i.e. gravity). Alternately within gravity mediation, supersymmetry could be
 broken by gaugino condensation\cite{Nilles:1982ik}. This mechanism is a likely 
 possibility within string theory.

 The strongest direct evidence supporting supergravity GUT models is the
 apparent experimental grand unification of the three gauge coupling constants
\cite{lang}.  This result is non-trivial not only because three lines do not
 ordinarily intersect at one point, but also because there is only a narrow
 acceptable window for $M_G$.  Thus one requires 
 $M_G{\tiny \begin{array}{l} >\\ \sim \end{array}} 5\times 10^{15}\ GeV$  
 so as not to violate current experimental bounds on proton decay
 for the $p\rightarrow e^+ +\pi^0$ channel (which occurs in almost all GUT
 models) and one requires $M_G\stackrel{<}{\sim} 5\times 10^{17}$ GeV $\cong
 M_{string}$ (the string scale) so that gravitational effects do not become
 large invalidating the analysis.   Further, assuming an MSSM type of particle
 spectrum between the electroweak scale $M_Z$ and $M_G$, acceptable grand
 unification occurs only with one pair of Higgs doublets and at most four
 generations.  Finally, naturalness requires that SUSY thresholds be at
 $M_S\stackrel{<}{\sim} 1$ TeV which turns out to be the case.  Thus the possibility
 of grand unification is tightly constrained.

 As discussed in Sec.(V) below,  the grand  unified models with R parity invariance produce
 a natural candidate for the dark matter observed astronomically. Further,  the amount of dark
 matter produced in the early universe can be calculated, and remarkably the theory naturally
 predicts a  relic density of dark matter  today of size  seen by WMAP and other observations.
 Thus SUGRA GUTS allows the construction of models valid from mass $M_G$ down 
 to the electroweak scale, and backwards in time to $\sim 10^{-8}$ sec after the Big Bang
 (when the dark matter was created),  a unification of particle physics and early universe
 cosmology.  
 At present, grand unification in SUGRA GUTs can be obtained to within about 2-3
 std. \cite{chank,das}  
 However, the closeness of $M_G$ to the Planck scale,
 $M_{P\ell}$ = $(\hbar c/8\pi G_N)^{1/2}\cong 2.4 \times 10^{18}$ GeV,
 suggests the
 possibility that there are O($M_G/M_{P\ell}$) corrections to these models.  One
 might, in fact, expect such structures to arise in string theory as
 nonrenormalizable operators (NROs) obtained upon integrating out the tower of
 Planck mass states.  Such terms would produce $\approx$ 1\% corrections at
 $M_G$
 which might grow to $\approx$ 5\% corrections at $M_Z$.  Indeed, as will
 be seen
 in  Sec.(II), it is just such NRO terms involving the hidden sector fields that
 give rise to the soft breaking masses, and so it would not be surprising
 to find
 such structures in the physical sector as well.  Thus SUGRA GUTs should be
 viewed as an effective theory and, as will be discussed in Sec.(VIII), with small
 deviations between theory and experiment perhaps opening a window to Planck
 scale physics.
 
 One of the fundamental aspects of the SM, not explained by that theory, is the
 origin of the spontaneous breaking of SU(2) x U(1).  SUGRA GUTS offers an
 explanation of this due to the existence of soft  SUSY 
 breaking masses at $M_G$.
 Thus as long as at least one of the soft breaking terms are present at $M_G$,
 breaking of SU(2) x U(1) can occur at a lower energy \cite{chams,inou},  
 providing
 a natural Higgs mechanism.  Further, radiative breaking occurs at the electroweak
 scale provided the top quark is heavy ie. 100 GeV
 $\stackrel{<}{\sim}m_t\stackrel{<}{\sim}$ 200 GeV.  The minimal SUGRA 
 model\cite{chams,Hall:1983iz,Nath:1983aw}(mSUGRA), which assumes universal soft
 breaking terms, requires only four
 additional parameters and one sign to describe all the interactions and masses
 of the 32 SUSY particles.  Thus the mSUGRA is  predictive model producing many 
 sum rules among the sparticle  masses \cite{Martin:1993ft},  and
 for that reason  the model is used in much of the phenomenological analysis of
 the past decades.  However, we will see in Sec.(II) that there are reasons to
 consider non-universal extensions of the mSUGRA, and inclusion of the nonuniversalities
 can produce significant modifications of the sparticle masses and their signatures. 
 
 \section{II. Soft Breaking Masses}\label{sec2}
 Supergravity interactions with chiral matter fields,
 $\lbrace\chi_i(x),\phi_i(x)\rbrace$ (where $\chi_i(x)$ are left (L) Weyl
 spinors and $\phi_i(x)$ are complex scalar fields) depend upon three functions of the
 scalar fields:  the superpotential $W(\phi_i)$, the gauge kinetic function
 $f_{\alpha\beta}(\phi_i,\phi_i^{\dag}$) (which enters in the Lagrangian as
 $f_{\alpha\beta}F^{\alpha}_{\mu\nu}F^{\mu\nu\beta}$ with $\alpha,\beta$ = gauge
 indices) and the Kahler potential $K(\phi_i,\phi_i^{\dag}$) (which appears in
 the scalar kinetic energy as
 $K_j^i\partial_{\mu}\phi_i\partial^{\mu}\phi_j^{\dag},$
 $K_j^i\equiv\partial^2 K^2 /\partial\phi_i\partial\phi_j^{\dag}$ and
 elsewhere).  W and K enter only in the combination
 \begin{equation}
 G(\phi_i,\phi_i^{\dag})=\kappa^2 K(\phi_i,\phi_i^+)+\ell n
 [\kappa^6\mid W(\phi_i)\mid^2]
 \label{1.1}
 \end{equation}
 where $\kappa = 1/M_{P\ell}$.  Writing
 $\lbrace\phi_i\rbrace=\lbrace\phi_a, z\rbrace$ where $\phi_a$ are physical
 sector fields (squarks, sleptons, higgs) and z are the hidden sector fields
 whose VEVs $\langle z\rangle ={\cal O}(M_{P\ell})$ break supersymmetry, one
 assumes that the superpotential decomposes into a physical and a hidden part,
 \begin{equation}
 W(\phi_i) = W_{phy}(\phi_a, \kappa z) +W_{hid}(z)
 \label{1.2}
 \end{equation}
 Supersymmetry breaking is scaled by requiring $\kappa^2 W_{hid} = {\cal
 O}(M_S)\tilde W_{hid}(\kappa z)$ and the gauge hierarchy is then
 guaranteed by the additive
 nature of the terms in Eq.(\ref{1.2}).  Thus only gravitational interactions remain to
 transmit SUSY breaking from the hidden sector to the physical sector.
 
 A priori, the functions W, K and $f_{\alpha\beta}$ are arbitrary.
 However, they
 are greatly constrained by the conditions that the model correctly reduce to
 the SM at low energies, and that non-renormalizable corrections be scaled
 by $\kappa$
 (as would be expected if they were the low energy residue of string physics of
 the Planck scale).  Thus one can expand these functions in polynomials of the
 physical fields $\phi_a$
  \beqn
 f_{\alpha\beta} (\phi_i) = c_{\alpha\beta} + \kappa
 d^a_{\alpha\beta}(x,y)\phi_a+\cdots,\nonumber\\
 W_{phys}(\phi_i)=\frac{1}{6}\lambda^{abc}(x)\phi_a\phi_b\phi_c+\frac{1}{24}
 \kappa\lambda^{abcd}(x) \phi_a\phi_b\phi_c\phi_d+\cdots,\nonumber\\
 K(\phi_i,\phi_i^{\dag})= \kappa^{-2} c(x,y)+c^a_b(x,y)\phi_a\phi_b^{\dag}\quad\quad~~~~~~~~~~~~~~~~~~~~~~~~~~\nonumber\\
 +
 (c^{ab}(x,y)\phi_a\phi_b + h.c.) +
  \kappa(c^a_{bc}\phi_a\phi_b^{\dag}\phi_c^{\dag} + h.c.)+\cdots. 
  \label{1.3}
 \end{eqnarray}
Here  x=$\kappa z$ and y =$\kappa z^{\dag}$, so that $\langle x\rangle$,
 $\langle y\rangle = {\cal O}$ (1).  The scaling hypothesis for the NRO's imply
 then that the VEVs of the coefficients $c_{\alpha\beta},~c_{\alpha\beta}^a,
 ~\lambda^{abc},~c,~c_b^a,~c^{ab}$ etc. are all ${\cal O}$(1).
  The holomorphic terms in K labeled by $c^{ab}$ can be transformed from K to W
 by a Kahler transformation, $K\rightarrow K-(c^{ab}\phi_a\phi_b + h.c.)$ and
 \begin{equation}
 W\rightarrow W exp [\kappa^2  c^{ab}~\phi_a\phi_b] = W +
 {\tilde\mu}^{ab}\phi_a\phi_b + \cdots
 \label{1.4}
 \end{equation}
 where $\tilde\mu^{ab}(x,y)=\kappa^2 Wc^{ab}$.  Hence  $\langle
 \tilde{\mu}^{ab}\rangle=\cal O (M_S)$, and one obtains a $\mu$-term with the 
 right order of magnitude after SUSY breaking provided only that $c^{ab}$ is not 
 zero \cite{soni}.
 The cubic terms in W are just the Yukawa couplings with
 $\langle\lambda^{abc}(x)\rangle$ being the Yukawa coupling constants.  Also
 $\langle c_{\alpha\beta}\rangle =\delta_{\alpha\beta}$, $\langle
 c_b^a\rangle=\delta_b^a$ and $\langle c_{xy}\rangle$ = 1 ($c_x\equiv\partial
 c/\partial x$ etc.) so that the field kinetic energies have canonical
 normalization.
 
 The breaking of SUSY in the hiddden sector leads to the generation of a series
 of soft breaking terms \cite{chams,Barbieri:1982eh,Hall:1983iz,Nath:1983aw,alte}.
 We consider here the case where $\langle
 x\rangle = \langle y\rangle$ (i.e. the hidden sector SUSY breaking does not
 generate any CP violation) and state the leading terms.  Gauginos gain a soft
 breaking mass term at $M_G$ of
 $\tilde m_{\alpha\beta}\lambda^{\alpha}\gamma^0\lambda^{\beta}$
 ($\lambda^{\alpha}$ = gaugino Majorana field) where
  \begin{equation}
 \tilde m_{\alpha\beta}= \kappa^{-2}\langle
 G^i (K^{-1})^i_j Re f_{\alpha\beta j}^{\dag}\rangle m_{3/2}
 \label{1.5}
 \end{equation}
 Here $G^i\equiv\partial G/\partial\phi_i$, $(K^{-1})^i_j$ is the matrix
 universe of $K_j^i$, $f_{\alpha\beta j} =\partial
 f_{\alpha\beta}/\partial\phi_j^{\dagger}$ and $m_{3/2}$ is the gravitino mass:
 $m_{3/2} =
 \kappa^{-1}\langle exp [G/2]\rangle$
   In terms of the expansion of Eq.(\ref{1.3})
 one finds 
 \beqn
 \tilde m_{\alpha\beta}= [c+\ell n (W_{hid})]_x Re~c_{\alpha\beta y}^{\ast} m_{3/2},
 \label{1.6}
 \eeqn 
 and $m_{3/2}$ = (exp $\frac{1}{2} c)\kappa^2 W_{hid}$ where it is
 understood from
 now on that x is to be replaced by its VEV in all functions (e.g.
 c(x)$\rightarrow c(\langle x\rangle) = {\cal O}(1))$ so that 
 $m_{3/2} = {\cal O}(M_S).$  
 One notes the following about Eq.(\ref{1.6}):  (i)  For a
 simple GUT group, gauge invariance implies that $c_{\alpha\beta}\sim
 \delta_{\alpha\beta}$ and so gaugino masses are universal (labeled by $m_{1/2}$)  
 at mass scales above
 $M_G$.  (ii)  While $m_{1/2}$ is scaled by $m_{3/2} = {\cal O}(M_S)$, it can
 differ from it by a significant amount.  
 (iii)  From Eq.(\ref{1.6}) one sees that it is
 the NRO such as $\kappa zm_{3/2}\lambda^{\alpha}\gamma^0\lambda^{\alpha}$ that
 gives rise to $m_{1/2}$.  Below $M_G$, where the GUT group is broken, the
 second term in the $f_{\alpha\beta}$ 
of  Eq.(\ref{1.3}) would also contribute yielding a NRO of size $\kappa
 d^a_{\alpha\beta}\phi_a m_{3/2}$ $\lambda^{\alpha}\gamma^0\lambda^{\beta}\sim
 (M_G/M_{P\ell})$ $m_{3/2} \lambda\gamma^0\lambda$ \cite{hill} for fields
 with VEV
 $\langle \phi_a \rangle = {\cal O} (M_G)$ which break the GUT group.  Such
 terms give small corrections to the universality of the gaugino masses and
 affect grand unification.  They are discussed in Sec.(VIII).
 
 The effective potential for the scalar components of chiral multiplets is given
 by \cite{chams,crem}
 \beqn
 V= e^{\kappa^2 K}~[( K^{-1})^j_i( W^i+\kappa^2 K^i
 W)~(W^j+\kappa^2 K^j W)^{\dag}
-3\kappa^2\mid W\mid^2] + V_D,\nonumber\\
 V_D = \frac{1}{2}
 g_{\alpha}g_{\beta}~(Ref^{-1})_{\alpha\beta}~(K^i(T^{\alpha})_{ij}\phi_j)~(
 K^k(T^{\beta})
 _{k\ell}\phi_{\ell}),~~~~~~~~~~~~~~~~~~~~~~~~
 \label{1.7}
 \eeqn
 where $W^i=\partial W/\partial\phi_i$ etc., and 
 where $g_{\alpha}$ are the gauge coupling constants.  Eqs.(\ref{1.2}-\ref{1.4}) then lead to
 the following soft breaking terms at $M_G$:
 \begin{equation}
 V_{soft} = ( m_0^2)^a_b~\phi_a\phi_b^{\dag}+\biggl[\frac{1}{3}
 {\tilde A}^{abc}\phi_a\phi_b\phi_c + \frac{1}{2}{\tilde
 B}^{ab}\phi_a\phi_b+h.c.\biggr]
 \label{1.8}
 \end{equation}
 In the following, we impose for simplicity 
  the condition that the cosmological constant vanish
 after SUSY breaking, i.e. $\langle V\rangle = 0$.
 [One of course  could accommodate the tiny cosmological constant suggested by the 
 supernova observation.] 
  This is a fine tuning of
 ${\cal O} (M_S^2M_{P\ell}^2)$.  
 From Eq.(\ref{1.7}) one notes  that the soft breaking terms are in general not universal
 unless one assumes that the fields z couple universally to the physical sector. 
  \section{III. Radiative Breaking and the Low Energy Theory}\label{sec3}
   In Sec.(II), the SUGRA GUT model above the GUT scale i.e. at $Q >M_G$ was
 discussed.  Below $M_G$ the GUT
 group is spontaneously broken, and we will assume here that the SM group,
 SU(3) x SU(2) xU(1), holds for $Q < M_G$.  
 Contact with accelerator
 physics at low
 energy can then be achieved using the renormalization group equations (RGE)\cite{Martin:1993zk} 
 running from $M_G$ to the electroweak scale $M_Z$.  As one proceeds
 downward from
 $M_G$, the coupling constants and masses evolve, and provided at least one soft
 breaking parameter and also the $\mu$ parameter at $M_G$ is not zero, the
 large top
 quark Yukawa can turn the $H_2$ running (mass)$^2$, $m_{H_{2}}^2(Q)$,
 negative at
 the electroweak scale \cite{inou}. Thus the spontaneous breaking of
 supersymmetry at
 $M_G$ triggers the spontaneous breaking of SU(2) xU(1) at the electroweak
 scale.
 In this fashion all the masses and coupling constants at the electroweak
 scale can
 be determined in terms of the fundamental parameters (Yukawa coupling constants
 and soft breaking parameters) at the GUT scale, and the theory can be
 subjected to
 experimental tests.
 
 The conditions for electroweak symmetry breaking arise from minimizing the
 effective potential V at the electroweak scale with respect to the Higgs VEVs
 $v_{1,2} = \langle H_{1,2}\rangle$.  This leads to the equations \cite{inou}
  \begin{equation}
 \mu^2 =\frac{\mu_1^2-\mu_2^2tan^2\beta}{tan^2\beta-1} - \frac{1}{2}
 M_Z^2;~~~sin^2\beta = \frac{-2B\mu}{2\mu^2+\mu_1^2+\mu^2_2}
 \label{1.9}
 \end{equation}
 where $tan\beta = v_2/v_1$, B is the quadratic soft breaking parameter
 ($V_{soft}^B=B\mu H_1H_2$),  $\mu_i=m_{H_{i}}^2+\Sigma_i$, and
 $\Sigma_i$ are loop corrections \cite{Arnowitt:1992qp}.
  All parameters are running parameters at the
 electroweak scale which one takes for convenience
 to be $Q\simeq \sqrt{m_{\tilde {t_1}} m_{\tilde {t_2}}}$ to minimize loop corrections.  Eq.(\ref{1.9}) then
 determines the $\mu$ parameter and allows the elimination of B in terms of
 $tan\beta$.  This determination of $\mu$ greatly enhances the predictive
 power of the model.
  
  In general there are two broad regions of electroweak symmetry breaking implied 
  by the  soft parameters appearing in Eq.(\ref{1.9}). One region is where the soft
  parameters can be arranged to  lie on the surface of an ellipsoid with their radii
  fixed by  the value of $\mu$. In this case  for fixed $\mu$, $m_0$ and $m_{\frac{1}{2}}$ 
  cannot get  too large  since the surface  of an ellipsoid is a closed surface. 
  However, it turns out that when loop corrections\cite{Arnowitt:1992qp}
   to the effective potential are included 
  the nature  of electroweak symmetry breaking can change  rather drastically. 
  Then the
  soft parameters  instead of lying on the surface of an ellipsoid,  lie on the surface
  of a hyperboloid and this branch may appropriately be called  the 
  hyperbolic branch (HB)(see the first paper of \cite{Chan:1997bi}).    
  Since the surface of  a hyperboloid in open, the soft parameters can get large  
  with $\mu$ fixed. Specifically,  the HB allows TeV size scalars with small values of $\mu$
  and thus small fine tunings. The region  of  TeV size  scalars  is also known 
  as the Focus Point (FP) region (see the second paper of \cite{Chan:1997bi}).

 The renormalization group equations evolve the universal gaugino mass
 $m_{1/2}$ at
 $M_G$ to separate masses for SU(3), SU(2) and U(1) at $M_Z$
 \begin{equation}
 \tilde m_i =(\alpha_i(M_Z)/\alpha_G)m_{1/2};~~~i = 1,2,3
 \label{1.10}
 \end{equation}
 where at 1-loop, the gluino mass $m_{\tilde g}=\tilde m_3$ \cite{mart}.
 The simplest model is the one with universal soft breaking masses.
  This model
 depends on only the four SUSY parameters and one 
 sign at the GUT scale\cite{chams,Nath:1983aw,Hall:1983iz} 
   \begin{equation}
 m_0, \quad m_{1/2}, \quad A_0, \quad B_0, \quad sign(\mu_0)
 \label{1.11}
 \end{equation}
 Alternately, at the electroweak scale one may choose
 $m_0\  $,  $m_{\frac{1}{2}}$, $A_t\  $, $tan\beta\  $, and sign($\mu$)
  as the independent parameters.
  Universality can be derived in a variety of ways. 
  From a string view point it could arise, for example, 
  from   dilaton dominance,  or when modular weights are all equal, 
  and in both GUTS and in strings it could arise  from a  family symmetry at the GUT/string scale.
 This case has been extensively discussed in the 
 literature 
 \cite{ross,Allanach:2000ii,Djouadi:2001yk,Roszkowski:2007va,Zerwas:2008ck}.
  The deviations from universality can be  significant, however, and affect 
 $\mu^2$ and sparticle masses, and these parameters play a
 crucial role in predictions of the theory.  
  Several analyses exist where the SUGRA models have been extended to include
  non-universalities \cite{Olechowski:1994gm,berez,Nath:1997qm,nonuni2}. These extensions include
  non-universalities  in the gaugino sector, in the Higgs sector and in the third generation
  sector consistent with flavor changing neutral current constraints. 

 \section{IV. Supersymmetric corrections to electroweak phenomena}\label{sec4}
\begin{figure}[h]
\includegraphics[width=11cm,height=9cm]{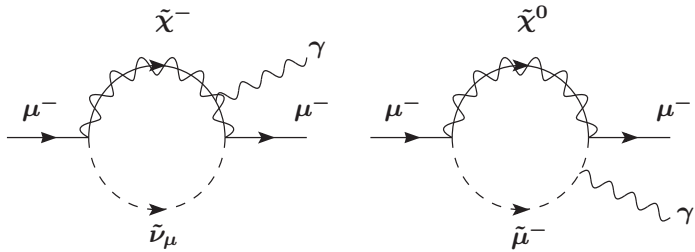}
\vspace{-2.5cm}
\caption{Supersymmetric electroweak contributions to $g_{\mu}-2$}
\label{fig1}
\end{figure}

 SUGRA models make  contributions to all electroweak processes at the loop level through the exchange
 of sparticles.  We discuss here some of the more prominent ones which include the 
 muon anomalous magnetic moment $g_{\mu}-2$, and the the flavor changing neutral current 
 processes $b\to s\gamma$ and $B^0_s\to \mu^+\mu^-$. 
 These processes are all probes  of new physics.
    Thus in    $g_{\mu}-2$   the sparticle
    loops at one loop make contributions  (see Fig.\ref{fig1})\cite{Yuan:1984ww}
     which are comparable to the electroweak
    contributions 
        from the Standard Model\cite{Yuan:1984ww}.
   The most recent evaluations of the difference between 
    experiment and theory give for $\delta a_{\mu}=(g_{\mu}^{exp}-g_{\mu}^{SM})/2$,
    the result\cite{Davier:2009zi}
        \beqn
    \delta a_{\mu}= (24.6\pm 8.0) \times 10^{-10}
    \label{1.12}
    \eeqn
     If the above  result holds up it would imply upper limits on sparticle masses
     within the range of the LHC energies
    \footnote{In most extra dimension models the corrections to $g_{\mu}-2$ 
    are rather small~\cite{Nath:1999aa} and it is difficult to accommodate a deviation of size Eq.(\ref{1.12}).}.     
     These conclusions were already drawn 
     earlier\cite{Yuan:1984ww,lopez,chatto,fm}.

   Flavor changing neutral current processes also provide
         an important constraint on supergravity unified models.
         A process of great interest here is the decay $b\rightarrow
         s+\gamma$.  Further, it is well known from the early days that 
         $b\rightarrow  s+\gamma$
          experiment imposes an important constraint on the parameter space  of 
          supergravity models\cite{Nath:1994tn} and specifically on the analysis of dark matter.         
           The current experimental value for this branching ratio from the
          the Heavy Flavor Averaging Group 
(HFAG)   \cite{Barberio:2008fa} along with the BABAR, 
Belle and CLEO  experimental results gives
$
{\mathcal Br}(B \to X_s \gamma) =(352\pm 23\pm 9) \times 10^{-6}.   
$
      In the SM this  decay proceeds
         at the loop level  with the exchange of W and Z bosons and the most recent evaluation
          including  the next to next leading order (NNLO) QCD corrections is given by \cite{Misiak:2006zs} 
$
{\mathcal Br}(b\rightarrow s\gamma) =(3.15\pm 0.23) \times 10^{-4}.
$
      In supersymmetry there are additional diagrams
           which contribute to this process \cite{bert}.
           Thus in SUGRA unification
           one has contributions from the exchange of  the charged Higgs,
           the charginos, the neutralinos and from the gluino.
           It is well known that the 
                   contribution from the charged Higgs exchange
           is always positive \cite{hewett}  while the contribution 
   from the exchange of the other SUSY particles  can be either positive or 
   negative with the
           contribution of the charginos being usually the dominant
           one \cite{diaz}.
          
           A comparison of the experimental and theoretical evaluations  in the SM 
             point to the possibility that a positive correction to the 
         SM value is needed. As noted above such a positive correction can arise
         from supersymmetry specifically from the exchange  of the charged higgs
         which implies the possibility of a relatively light charged Higgs. 
          Further, over most of the parameter space the chargino exchange
         contributions are often negative  pointing to a cancellation between the charged
         Higgs and the chargino exchange contributions and also hint at the possibility
         of a relatively light chargino and possibly of a relatively light stop.

The rare process  $B_s\to \mu^+\mu^-$  is of interest  as it is a probe of  phyiscs 
beyond the standard model\cite{Babu:1999hn,Bobeth:2001sq}.
The branching ratio for this process in the SM is 
${\mathcal Br}( B_s \to \mu^{+}\mu^{-})=(3.1\pm 1.4) \times 10^{-9}$ 
(for $V_{ts}=0.04\pm 0.002)$). In supersymmetric models it can get large for large
$\tan\beta$ since decay branching ratio increases as $\tan^6\beta$.  The current
experimental limit at  95\% (90\%) C.L. reported  by CDF is 
${\mathcal Br}( B_s \to \mu^{+}\mu^{-})=5.8 \times 10^{-8}$ ($4.7 \times 10^{-8}$)  \cite{2007kv}. 
Since in supersymmetric theories this branching ratio can increase as $\tan^6\beta$ the 
experimental data does  constrain the analysis at least for large $\tan\beta$
and the implications of  this constraint have been analyzed in several 
works\cite{Dedes:2002zx,Arnowitt:2002cq}.
Additionally, this specific decay is very sensitive to CP phases and thus 
the experiment also 
constrains the CP phases in SUGRA  models in certain regions of the parameter 
space\cite{Ibrahim:2002fx}. 
  \section{V. Dark matter in SUGRA unification}\label{sec5} 
    As mentioned earlier one of the remarkable results of
    supergravity grand unification with R parity invariance is the
    prediction that  the  lightest neutralino $\chi_1^0$ is the LSP over 
    most of the
    parameter space ~\cite{lsp}.  In this part of the
    parameter space  
     the $\chi_1^0$ is a
    candidate for cold dark matter (CDM).
     We discuss now the relic density of $\chi_1$ within the framework
     of the Big Bang Cosmology. The quantity that is computed theoretically
     is $\Omega_{\chi_1} h^2$ where  $\Omega_{\chi_1}=\rho_{\chi_1}/\rho_c$,
      $\rho_{\chi_1}$ is the neutralino relic density and $\rho_c$
     is the critical relic density needed to close the universe,
    $\rho_c\ =\ 3H^{2 }/8\pi G_N$, and $H=h 100 km/s Mpc$ 
   is the Hubble 
    constant. 
             One of the important elements in the computation of the relic
         density concerns the correct thermal averaging of the quantity
         ($\sigma v$) where  $\sigma$ is the neutralino  annihilation
         cross section in the early universe and 
    $v$ is the relative neutralino velocity.     Normally
         the thermal average is calculated by first making the approximation
         $\sigma v=a+b v^2$ and then evaluating its thermal
         average \cite{kolb}.
         However, it is known that such an approximation breaks down
         in the vicinity of thresholds and poles \cite{greist}.
         Precisely such a
         situation exists for the case of the annihilation of the
         neutralino through the Z and Higgs poles. An accurate analysis
         of the neutralino relic density in the presence of Z and Higgs
         poles was given in ref. \cite{dark1} 
        \footnote{The  analysis of \cite{dark1} has been used to show that 
annihilation near a Breit-Wigner pole generates a significant enhancement of 
$<\sigma v>_H$ in the halo of the galaxy relative to $<\sigma v>_{X_f}$
at the freezeout\cite{Feldman:2008xs}.}
          and  similar analyses  have
         also been carried out since  by other authors \cite{Baer:1995nc}.
          There are a number of possibilites for  the  detection of
           dark matter both direct and indirect \cite{jungman}.
           We discuss first
            the
           direct method which involves the scattering of incident neutralino
           dark matter in the Milky Way from nuclei in terrestial targets. 
    The event
           rates
        consist of two parts \cite{goodman}: one involves an axial interaction
           and the other a scalar interaction. The axial
           (spin dependent) part $R_{SD}$
           falls off as $R_{SD}\sim 1/M_N$ for large $M_N$ where
           $M_N$ is the mass of the target nucleus, while
           the scalar (spin independent) part behaves as
           $R_{SI}\sim M_N$ and increases with $M_N$. Thus for heavy
           target nuclei the spin
           independent part $R_{SI}$ dominates over most of the
           parameter space of the model. 
          
             In recent years the direct detection dark matter experiments have 
             begun to provide significant bounds on  
             the spin independent neutralino -proton 
             cross section $\sigma_{\tilde\chi^0 p}$ and thus theoretical computations
             of this quantity can be  directly compared with the data. The predictions of the
             SUGRA models lie over a wide range. With typical asssumptions of naturalness
             on sparticle masses below 1 TeV,  $\sigma_{\tilde\chi^0 p}$ can lie in the range
             $10^{-43}$ cm$^2$ to $10^{-48}$ cm$^2$.  The current experiments
           such as CDMS\cite{Ahmed:2008eu}           
            and  XENON\cite{Angle:2007uj} have  already 
             begun to constrain a part of the SUGRA parameter space and improved
             experiments\cite{Projections} are expected to probe the  parameter space further\footnote{Recently the CDMS experiment has observed two 
             events fitting the behavior of WIMPS with a background of
             $0.6\pm 0.1$ events\cite{Ahmed:2009zw}. Further, data is needed to confirm this.}.        
             Inclusion of non-universalities is seen to
           produce definite signatures in the event rate
           analysis \cite{Nath:1997qm,berez}.
           The satisfaction of the relic density in SUGRA models can occur  via coannihilations.
            One of the most studied coannihilation is with the coannihilation of the neutralino with the
             stau.                 However, 
                    with the inclusion of non-universalities of soft parameters many other coannihilaitons become
                    possible such as coannihilations of the LSP with charginos, stops, gluinos, and
                    heavier neutralinos etc. The coannihilation with the gluino which occurs most dominantly
                    when the gluino is the NLSP and 
                    exhibits the interesting phenomenon that the sparticle production
                    cross sections are dominated by the gluino production making the observation of other
                    sparticles challenging\cite{Feldman:2009zc}. 
 
  In addition to  the direct detection dark matter experiments, there are also indirect signatures
  for dark matter. Thus, e.g., neutrinos arising from the annihilation of neutralinos in the center 
  of the Earth and Sun can produce detectable signals. Further, it was suggested  
   quite  sometime ago~\cite{Turner:1989kg} that the annihilation process
$\tilde\chi^0\tilde\chi^0\to W^+W^-$ with the subsequent decays of the W's, e.g., $W^+\to e^+\nu$ 
could generate a detectable  positron excess in anti-matter probes. One of the typical
problems encountered  in most theoretical analyses of the positron excess is the following: in order
to have the appropriate positron signal in PAMELA~\cite{pamela}  one needs to have a $\chi^0\chi^0$  annihilation 
cross section to $WW$ with  $<\sigma v>_{WW} \simeq 10^{-24} cm^3/s$. However, the relic
density has an inverse proportionality to the annihilation cross section at the freeze out, i.e.,
$\Omega_{\tilde \chi^0} h^2 \propto [\int_{x_f}^{\infty} <\sigma_{eff} v> \frac{dx}{x^2}]^{-1}$
which leads to too low a relic density. To overcome this problem most works  typically resort 
to large so called boost  factors. Effectively, what this implies is  that  the annihilation cross section
of dark matter is taken to satisfy the relic density and then to get the right strength positron 
signal a boost factor is  assumed. It  is argued that  such boost  factors can arise
from clumping of dark matter in the galaxy. However, while a boost factor of O(2-10) could arise
arise from clumping of dark matter in the galaxy, it appears unreasonable to assume 
large clumping factors (sometimes as large as  $10^3$ or even larger) to fit the data. 
In the context of the minimal supergravity model, 
one simple solution arises due to the automatic suppression of 
the relic density from coannihilation effects with hidden sector matter in SUGRA models
with an extended $U(1)^n$ sector\cite{Feldman:2009wv}. 
In this case  with n=3, one finds  good fits to the positron
excess from PAMELA while maintaining the neutralino relic  density in the WMAP~\cite{WMAP}
 error  corridor. At the same time one can maintain  compatibility with the  anti-proton flux~\cite{pamela}
and the photon flux from the FERMI-LAT experiment~\cite{Abdo:2009zk}.

     \section{VI. Signatures at colliders}\label{sec6}
Sparticle  decays produce  missing energy signals since 
   at least  one of the carriers of missing energy will be the
  neutralino. Signals of this type were studied early on after the
  advent of supergravity models in the supersymmetric decays of the
  W and Z bosons \cite{swein} and such analyses have since been extended
  to the decays of all of the supersymmetric particles (For a review of sparticle 
  decays see \cite{applied}).
   Using these
  decay patterns one finds  a variety of  supersymmetric signals for SUSY
  particles at colliders where SUSY particles are expected to be
  pair produced when sufficient energies in the center of mass system
  are achieved.  One signal of special interest in the search for
  supersymmetry is the trileptonic signal through 
  off shell $W^*$ production as well as via other production and decay chains   
   \cite{trilep}.
  For example in $p\bar p$
  collisions one can have $p\bar p\rightarrow \tilde \chi^{\pm}_1+
  \tilde \chi^0_2+X
  \rightarrow(l_1\bar \nu_1\tilde \chi^0_1)+(l_2\bar l_2\tilde \chi^0_1)+X$ which
  gives a signal of three leptons and missing energy. 
         In addition to the trileptonic signal there is  a long list of possible signatures for the
         discovery of supersymmetry and test of SUGRA models. These include multileptons
         and multijet and  missing energy. Thus  one can devise a variety of combinations
         with n number of leptons ($e$ or $\mu$), m number of $\tau$'s, k number of jets 
         $(m,n=0,1,2,3,..; k\geq 2)$ leading to a large number of possibilities. Further, 
         one can add to this list tagged b jet signals and kinematical signatures
         such as   missing transverse
  momentum $P_T^{miss}$, effective mass $P_T^{miss}  +\sum_j P_T^j$,
  invariant mass of $e^+e^-$, $\mu^+\mu^-$, $\tau^+\tau^-$, and invariant mass of all jets
  which provide important signatures.  
  
   An important  result  concerns the fact that one can utilize measurements at the
   LHC to predict phenomena related to dark matter, showing the unification of particle
   physics and cosmology.  Within the mSUGRA framework, existing constraints on the 
   parameter space combined with the cold dark matter constraints pick out three regions:
   (i) the $\tilde\chi_1-\tilde \tau_1$ coannihilation (CA) region where $m_0$ is small but
   $m_{1/2}$ can rise to 1 TeV, (ii) the hyperbolic (HB)/focus (FP) region, where $m_{1/2}$ 
   is relatively small but $m_0$  is large, and (iii) the pole region\cite{lsp} (alternately called the
   funnel region) 
   where annihilation in the early universe goes via heavy Higgs poles. 
   For the CA region it can be shown\cite{Arnowitt:2008bz} 
  that purely from the measurements at the LHC one 
   can predict the dark matter 
   relic density with an uncertainty of $6\%$ with 30fb$^{-1}$ of data
    which is comparable to the uncertainty in the determination of the relic density by WMAP. 
   The relevant signal here consists of low energy $\tau$ leptons from $\tilde \chi_2^0\to \tau\tilde \tau_1
   \to \tau\tau\tilde\chi_1^0$ where the mass difference of the $\tilde \tau_1$ and $\tilde \chi_1^0$ is
   constrained to lie within in 5-15 GeV  by the current experimental bounds\cite{Arnowitt:2007nt}.
   In addition it is possible to test experimentally the universality of the gaugino masses  and if not,
   measure the amount of non-universality as well as obtain precision measurements of the gaugino
   mass, squark and lighter stau masses. A similar analysis may be carried out in the HB/FP  
   region\cite{DK2009} and very likely can be done for the ILC\cite{Khotilovich:2005gb}.

 We discuss now an  approach by which the LHC data can be used  to discriminate among
 a variety of models. This approach utilizes the idea of sparicle mass hierarchies which we
 now describe. Thus as mentioned already in  
           MSSM there are 32 supersymmetric particles including the Higgs fields.  In general 
they can generate a large number of mass hierarchies.  Assuming the lightest sparticle is the 
lightest neutralino, there are still in excess of $10^{25}$ possible mass hierarchies in which 
the sparticles can arrange themselves. 
Of these  only one will eventually  be realized if all the  sparticle masses are  finally
measured at the LHC or  in other future collider experiments. The question then is how predictive are
SUGRA models in pinning down the mass hierarchical patterns.  
The above  question can be answered  within the SUGRA framework including the REWSB constraints, 
the WMAP and other relevant experimental constraints\cite{Feldman:2007zn}.  An analysis along these lines but limited
to four particle mass hierarchies aside from the lightest Higgs boson mass would in general lead to
roughly $O(10^4)$ such mass hierarchical patterns. However, within the mSUGRA framework 
with the constraints mentioned  above one finds that  the number of possibilities reduces to just 16 
for $\mu$ positive and 6 more for  $\mu$ negative. These possibilities  are  labeled as minimal supergravity patterns 
mSP1-mSP16 for $\mu>0$ and mSP17-mSP22 for $\mu<0$.  
These allowed set of models can be further subdivided into classes according to their next to the lowest
mass particle (NLSP).  Thus one finds the dominant sub classes of patterns among mSP1-mSP16 to be
the Chargino Pattern, Stau Pattern, 
Stop Pattern  and Higgs Pattern. In addition for $\mu<0$ one finds additional Stau and Stop 
Patterns and also a Neutralino Pattern where the second  neutralino is the NLSP\cite{Feldman:2007zn}.

 A similar analysis can also be carried out for the non-universal SUGRA case. Here allowing for 
 non-universalities in the  Higgs  sector, gaugino sector and in the third generation sector
  one finds 22 new sparticle patterns for the first four sparticles (excluding the lightest Higgs boson).
 These are labeled NUSP1-NUSP22. It is found that no new patterns arise from non-universalities
 in the Higgs sector and all the new patterns are from non-universalties in the gaugino sector and 
 in the third generation sectors. It is shown in \cite{Feldman:2007zn} that these new patterns have
 distinctive signatures and can be discriminated by appropriate choices of events with leptons, jets
 and missing energy.   
  Specifically
  using leptons, jets and missing energy events one can discriminate between the stau 
  coannihilation branch and the hyperbolic branch/focus point region. Thus it is also found 
  that one can identify the origin of dark matter using LHC data.  

\section{VII. CP Violation}\label{sec7} 
The minimal supergravity model with universal soft breaking has two independent CP violating
phases which can be chosen to be the phase of the $\mu$ parameter ($\theta_{\mu}$) and the
phase of the  trilinear coupling $A_0$ ($\alpha_{A_0}$).  In the more general  soft breaking of
the non-universal supergravity model, one may have many  more phases. For instance,
with non-universal gaugino masses each of the gaugino masses in the $U(1)_Y$, $SU(2)_L$
and $SU(3)_C$ may have a phase, i.e., $\tilde m_i =|\tilde m_i| e^{i\xi_i}$ (i=1,2, 3)
of which two are independent.
Similarly, the trilinear couplings may be complex and flavor dependent, so that $A_a = |A_a| e^{i\alpha_{A_a}}$.
For the most general allowed soft breaking in MSSM, the list of allowed phases is much larger,
and even after field redefinitions many CP phases remain. 
In general the CP phases lead to large  supersymmetric contributions to the electric  dipole moments (EDMs) of the
neutron and of the leptons leading to their EDMs far in excess of experiment.  These EDMs can be 
made compatible by a variety of means, such as through choice of small CP 
phases \cite{Weinberg:1989dx,Arnowitt:1990eh}, large sparticle  
masses \cite{na},  via the cancellation mechanism\cite{incancel2,aad}, or via the CP phases  arising only in the 
third generation\cite{Chang:1998uc}. If the CP phases are large they will lead to 
mixings ~\cite{Pilaftsis:1998dd} of CP even Higgs  
and CP odd Higgs states and there would be many phenomenological implications at colliders and
elsewhere~\cite{Ellis:2004fs,Ibrahim:2007fb}.  
  \section{VIII. Planck scale corrections and further tests of  SUGRA  GUT and Post GUT physics}\label{sec8}                    Because of the proximity of the GUT scale to the Planck scale
              one can expect corrections of size O($M_G/M_{Pl})$ to grand unification
                            where 
              $M_{Pl}$ is the Planck mass.
               For example, Planck scale corrections can modify the
               gauge kinetic energy function so that one has for the
               gauge kinetic energy  term -(1/4)$f_{\alpha\beta}
               F_{\alpha}^{\mu\nu}F_{\beta\mu\nu}$. 
                 For the minimal SU(5) theory, $f_{\alpha\beta}$ in SUGRA models 
               can assume the form
                $f_{\alpha\beta}=\delta_{\alpha\beta}+(c /2M_{Pl})
                d_{\alpha\beta\gamma}\Sigma^{\gamma}$ where
                $\Sigma$ is the scalar field in the 24 plet of SU(5).
               After the spontaneous breaking of SU(5) and a re-diagonalization
               of the gauge kinetic energy function, one finds a splitting
               of the $SU(3)\times SU(2)\times U(1)$ gauge coupling
               constants at the GUT scale. These splittings 
               generate a corrections to $\alpha_i(M_{Z})$, and 
                using the LEP data one can put constraints
               on c. One finds that \cite{das}$-1\leq c \leq 3$. 
          The Planck scale
               correction also helps relax the stringent constraint
               on $tan\beta$ imposed by $b-\tau$ unification. Thus in the
               absence of Planck scale correction one has that $b-\tau$
               unification requires $tan\beta$ to lie in two rather
               sharply defined corridors \cite{bbo}. One of these corresponds
               to a small value of $tan\beta$, i.e., $tan\beta \sim 2$
                 and the second a large value $tan\beta \sim 50$. This
                 stringent constraint is somewhat relaxed by the
                 inclusion of Planck scale corrections \cite{das}.

             SUSY grand unified models contain many sources of
             proton instability. Thus in addition to the
             p decay occuring via the exchange of superheavy  vector
             lepto-quarks,
             one has the possibility of p dacay from dimension (dim)
              4 (dim 3 in the superpotential) and dim 5
              (dim 4 in the superpotential) operators \cite{wein}.
              The lepto-quarks exchange would
              produce  $p\rightarrow e^+\pi^0$
              as its dominant mode with an expected lifetime \cite{gm} of
               $\sim 1\times
              10^{35\pm 1}{[M_X/10^{16}] }^{4}y $ where 
              $M_X \cong 1.1 \times 10^{16}$
               while the current lower limit on this decay mode from Super Kamiokande
                is $\sim 2\times 10^{33}$ yr.
              Thus the $e^+\pi^0$ mode may be at the edge of being accessible
              in  proposed experiments\cite{Rubbia:2009md} such as at DUSEL which will have improved
              sensitivities for this decay mode. 
                 Proton decay from dim 4 operators is much too rapid but is
                 easily forbidden by the imposition of R parity invariance.
                  The p decay from dim  5 operators is more involved.
                  It depends on
                  both the GUT physics as well as on the
                  low energy physics such as
                  the masses of the squarks and of the gauginos.
                                    Analysis in supergravity unified models\cite{Arnowitt:1985iy} shows that
                   one can make concrete predictions of the p decay
                   modes within these models once the sparticle spectrum is determined.                   
                   
                   Precision determination of soft SUSY breaking parameters
                   can be utilized as a vehicle for the test of the
                  predictions of supergravity grand unification.
                  Specifically it has been 
                  proposed that
                  precision measurement of the soft breaking parameters can also
                  act as a test of physics at the post GUT and string
                  scales\cite{Arnowitt:1997ui}. Thus, for example,
                  if one has a concrete
                  model of the soft breaking parameters at the string scale
                  then these parameters can be evolved  down to the
                  grand unification scale leading to a predicted set of
                  non-universalities there. If the SUSY particle spectra and
                  their interactions are known with precision at the
                  electro-weak scale, then this data can be utilized
                  to test a specific  model at the post GUT or string
                  scales. Future colliders such as the LHC \cite{baer}
                  and the NLC \cite{tsukamoto}
                   will allow one to make
                  mass measurements with significant accuracy. Thus
                  accuracies  of up to a few percent
                  in the mass measurements will be possible at these 
                  colliders allowing
                  a test of post GUT and string physics up to an
                  accuracy of $\sim 10\%$\cite{Arnowitt:1997ui}.

\section{IX. Conclusion}\label{sec9}
  Supergravity grand unification provides a framework for the
  supersymmetric unification of the electro-weak and the strong
  interactions where supersymmetry is broken spontaneously by
  a super Higgs effect in the hidden sector
   and the breaking communicated to the
  visible sector via gravitational interactions. The minimal version
  of the model based on a generation independent Kahler potential
  contains only four additional arbitrary parameters and one 
  sign in terms of which all the
  SUSY mass spectrum and all the SUSY interaction structure
   is determined. This model is thus very predictive. A brief summary
   of the  predictions and the phenomenological implications
   of the model were given. Many of the predictions of the model can be
   tested at current collider energies and at energies that would
   be achievable at the LHC. 
   We also discussed  here extensions of
   the minimal supergravity model to include non-unversalities in the
   soft SUSY breaking parameters. Some of the implications of these
   non-universalities on predictions of the model were discussed.
   Future experiments should be able  to see if  the predictions of
   supergravity unification are indeed verified in nature.




\begin{thebibliography}{999}
\bibitem{alte}
 Alternate possibilities are considered elsewhere in this book.
\end{thebibliography}
\end{document}